\documentclass[18pt,fleqn]{article}
\usepackage{amsmath}
 \usepackage{paralist}
  \usepackage{graphics}
   \usepackage{epsfig}
    \usepackage[colorlinks=true]{hyperref}
    \usepackage{amstext,amsfonts,amssymb,amsthm,amsmath,mathrsfs,epic,eepic}
\usepackage{subfigure}  
 \hypersetup{urlcolor=blue, citecolor=red}

\theoremstyle{definition}

\renewcommand{\eqref}[1]{Eq.~(\ref{#1})}

\newcommand{\figref}[1]{Fig.~\ref{#1}}

\newcommand{\be}{\begin{equation}}
\newcommand{\ee}{\end{equation}}
\newcommand{\secref}[1]{Section \ref{#1}}

\title
      {Adaptive response and enlargement of dynamic range}

\begin{document}
\maketitle

\centerline{\scshape Tamar Friedlander}
\medskip
{\footnotesize
 \centerline{Physics of Complex Systems}
   \centerline{Weizmann Institute of Science,
 Rehovot 76100, Israel.}
} 

\medskip

\centerline{\scshape Naama Brenner }
\medskip
{\footnotesize
 \centerline{Department of Chemical Engineering and Laboratory of Network Biology Research, }
   \centerline{Technion - Israel Institute of Technology, Haifa 32000, Israel.}
} %

\bigskip


\begin{abstract}
Many membrane channels and receptors exhibit adaptive, or desensitized,
response to a strong sustained input stimulus, often supported by protein activity-dependent inactivation.
Adaptive response is thought to be related to various cellular functions
such as homeostasis and enlargement of dynamic range by background compensation.

Here we study the quantitative relation between adaptive response and background compensation within a modeling
framework.
We show that any particular type of adaptive response is neither sufficient nor necessary for adaptive
enlargement of dynamic range. In particular a precise adaptive response, where system activity is maintained
at a constant level at steady state, does not ensure a large dynamic range neither in input signal nor in system
output. A general mechanism for input dynamic range enlargement can come about from the activity-dependent
modulation of protein responsiveness by multiple biochemical modification, regardless of the type of adaptive
response it induces. Therefore hierarchical biochemical processes such as methylation and phosphorylation
are natural candidates to induce this property in signaling systems.
\end{abstract}

\section{Introduction}
Organisms in general and cells in particular face the continuous challenge of sensing their environment and responding accordingly \cite{Kholodenko06,SensoryAdaptation07}. Cell membranes are populated by a variety of proteins, such as ligand-binding receptors and various signal responding channels, that carry out this sensing and transmit information to downstream processes in the cell. In many cases these sensing molecules can be described as having two distinct functional states, such as active/inactive for a receptor or conducting/non-conducting for a channel. The transitions between functional protein states often depend directly on incoming signals, but are also regulated on longer timescales by various cellular processes.

Many of these sensing and signaling proteins exhibit adaptive response, or desensitization,
following a strong and persistent stimulation.
The qualitative hallmark of such responses is
observed when exposing the system to a
step input signal: an abrupt change of response is followed by a slow relaxation on a longer time scale. These responses have been described and studied for
many years in different areas of biology. Recently they have been the topic of much theoretical work, mainly along two lines of research. On one hand, elaborate models were developed to describe specific systems (most notably bacterial chemotaxis \cite{Keymer06,MelloTu07,Hansen2008,Tindall08}, but also other systems). These models are highly successful in reconstructing experimental results, but due to their
complexity it often
remains unclear what model ingredient is responsible for which system property. This situation impedes on our
ability to generalize models from one system to the other and to distinguish universal from system-specific properties. The other line of research has taken a more abstract approach, characterizing and classifying small simple sub-networks with respect to their adaptive response properties \cite{Tyson03,Behar07,Francois08,MaTang09}.

An adaptive response in a signaling system indicates underlying processes on more than one timescale,
but does not necessarily imply any particular functionality. If the adaptive response is precise,
namely it returns at steady state to a fixed value insensitive to the input, it can serve a homeostasis mechanism
that keeps some variables constant in the face of changing environments. However, from a more general perspective
of the cell sensing its environment and transmitting information, it is not a-priori clear how adaptive response
is related to signal processing properties, for example the determination of effective input dynamic range.

Cellular signaling systems are known for the remarkably broad dynamic range of inputs to which they can respond. In
particular they can respond to relatively small changes in signal on top of large constant backgrounds, effectively
compensating the background and remaining sensitive to fluctuations around it. For example, the signaling system responsible for bacterial chemotaxis maintains
sensitivity to changes in nutrient gradients over 5 orders of magnitudes of ligand concentrations~\cite{Bourret91}.
Photoreceptors, as another example, respond to light intensities spanning 11 orders of magnitude \cite{PughLamb}.

The response to a broad dynamic range of input achieved by cellular signaling
systems was often associated with their adaptive response,
led by the intuition
that the return of the system to its previous state
allows renewed sensitivity to signal
\cite{BarkaiLeibler97,Hansen2008,MaTang09}.
However, the quantitative relation between the two properties was not examined in detail.
In this work we focus on this relation by utilizing a previously developed general model for adaptive response~\cite{FriedlanderBrenner09},
which has proven useful in making the distinction between universal and system-specific features.
We find that the relation between the two properties - adaptive response and dynamic range - is more subtle than may be expected; in particular, any form of adaptive response is neither sufficient nor necessary for implementation of efficient dynamic range enlargement by background compensation. In \secref{sec:BasicModel} we review for completeness a general class of 3-state models \cite{FriedlanderBrenner09} and its properties. In \secref{sec:DynRange_BasicModel} we quantify and compute the degree of dynamic range enlargement by these models, and conclude that state-dependent inactivation or any particular form of adaptive response is insufficient to ensure such an enlargement. In \secref{sec:ExtModel} we generalize the concept of protein unavailability (or inactivation) from a binary to a graded scale
of protein responsiveness,
and show that such a generalization robustly results in adaptive enlargement of the dynamic range, regardless of kinetic details or of the temporal form of adaptive response. We conclude by relating the results to experiments and reviewing some open questions.

\section{A three-state model for adaptive response}
\label{sec:BasicModel}
\begin{figure}[h]
\begin{center}
\includegraphics[width=0.6\textwidth]{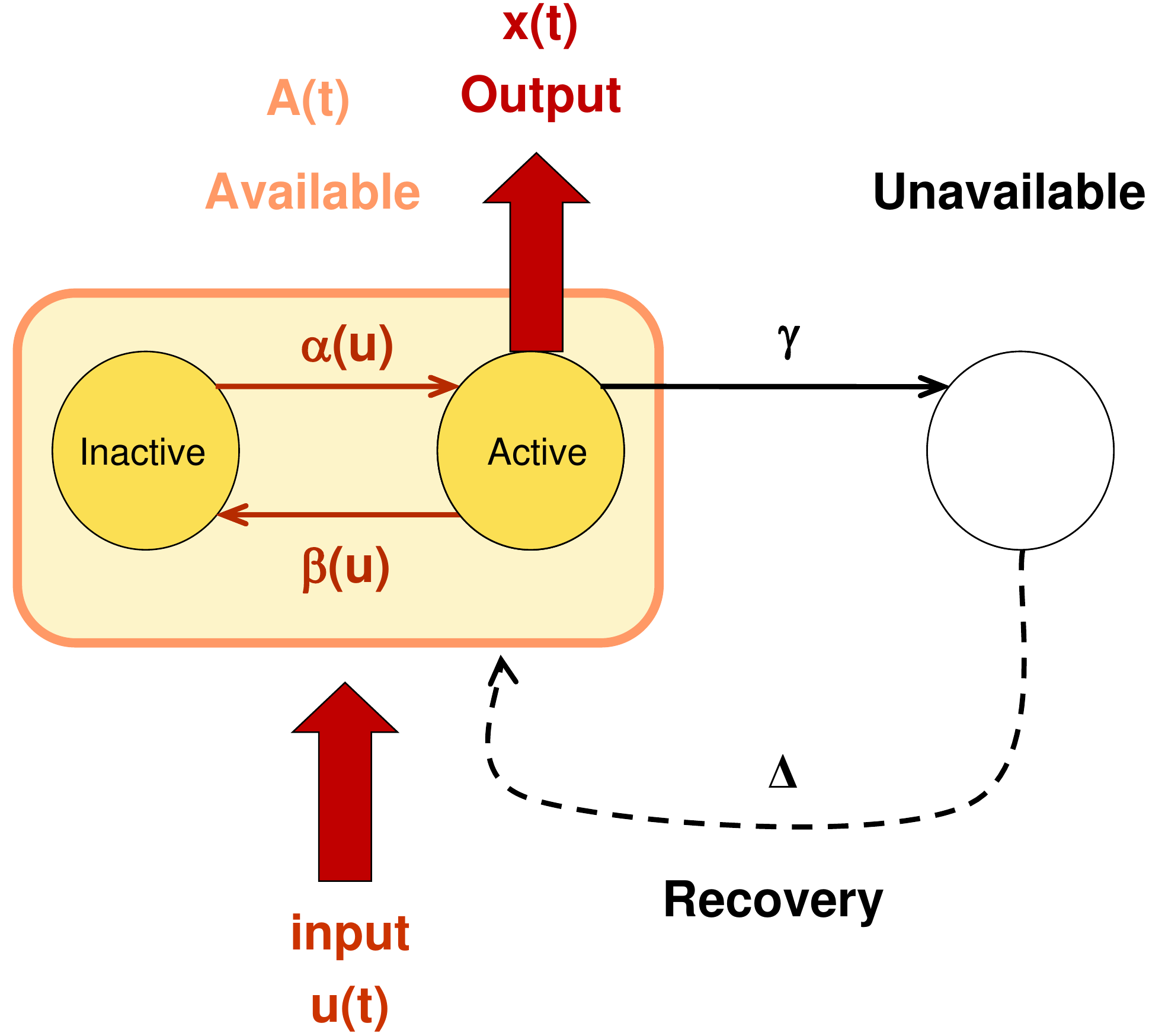}
\end{center}
      \caption{\label{fig:3-state}
General 3-state kinetic scheme providing an abstract model for state-dependent (i.e. activity-dependent) inactivation. Proteins can be active or inactive, with a balance between them depending on the input $u(t)$. Population of the active state $x(t)$, which equals the total activity, is the system's output. In addition proteins can transit, slowly and only through the active state, to an ``unavailable" state in which they cannot respond to the input. Kinetics of recovery from unavailability, denoted by the abstract term $\Delta$, may vary according to physical implementation found in various biological systems.
For detailed analysis see \cite{FriedlanderBrenner09}}
      \end{figure}

In our previous work we introduced a general simplified model for adaptive response unifying many biological
systems and enabling a mapping to control circuits \cite{FriedlanderBrenner09}. The
model is based on an ensemble of protein molecules that can be active or inactive, and in addition display slow
transition between available and an unavailable
pools. Available proteins can rapidly respond to the input signal, switching between the two distinct functional states active/inactive.
Unavailable proteins, by contrast,
cannot respond on a short time-scale but only after recovering back to the available pool. The physical mechanisms for proteins to become unavailable or to recover
back to the available pool are diverse; they share the properties of being slow and activity-dependent. The general structure of the model is described by
the following scheme:
\be
\underbrace{\mbox{inactive} \mathop{\rightleftarrows}_{\beta(u(t))}^{\alpha(u(t))}
\mbox{active x(t)}}_{\mbox{available A(t)}} \mathop{\rightleftarrows}_{\Delta}^{\gamma}
\mbox{unavailable}.
\label{eq:COI_scheme}
\ee

\noindent in which $u(t)$ is the input signal; $\alpha(u(t)),\beta(u(t))$ are the input-dependent transition
rates between active and inactive states of the protein; $\gamma$ is the rate constant for inactivation,
assumed to be first-order in the concentration of active molecules $x(t)$; and $\Delta$ is a general term for
recovery to availability that can be constant, first-order, or history-dependent (see also illustration in
\figref{fig:3-state}).
In this model the available pool of proteins, represented by the dynamical variable $A(t)$,
registers the system past activity and feeds it back to the observable system output $x(t)$
via multiplicative feedback. The equations describing the system dynamics are written in terms of the activity
$x(t)$ and availability $A(t)$:

\begin{eqnarray}
\label{eq:2D_diff_eq}
\dot{x}&=& \alpha(u(t))(A(t)-x(t))-\beta(u(t)) x(t)-\gamma x(t)+\Delta
\\\nonumber \dot{A}  &=& - \gamma x(t)+\Delta .
\end{eqnarray}

The direct influence of the input signal on the system enters through the transition rates, $\alpha(u),\beta(u)$. Assuming these transitions are rapid, each value of input $u$ instantaneously defines a balance between the two functional states, such that the activity reflects the input through the input/output function $p(u)$:
\be
\label{eq:adiab}
x = A\frac{\alpha(u)}{\alpha(u)+\beta(u)}=Ap(u),
\ee
\noindent where $A$ denotes the total concentration of available proteins. The details of this function depend
on the physical properties of the sensing proteins.
On longer timescales, the slow dynamics $A(t)$ comes into play. These dynamics reflect various cellular processes
that modulate the number of available proteins, and the details of these dynamics depend on the particular
mechanism of inactivation and recovery. We have shown in our
previous work that the dynamical variable $A(t)$ effectively encodes an integral over past system activity, with a kernel that depends on details of its kinetics. In particular, for different recovery kinetics $\Delta$ the induced adaptive response to step changes can take various temporal forms: exponential or power-law, precise or input-dependent. However, regardless of these forms, under conditions of timescale separation the response is still $x(t)\approx A(t)p(u(t))$.
Thus the response to abrupt changes will be characterized by a rapid change in $p(u)$ followed by slow changes in $A$ which multiplies $p(u)$.

The three-state general model presented above is equivalent to previously studied models in several special cases.
For a first-order recovery with conservation of the number of molecules, $\Delta = \delta (1-A)$, it is equivalent
to a model proposed and studied to describe the kinetics of voltage-gated ion channels \cite{Marom94}.
This system exhibits exponential adaptive response which relaxes to a steady-state level dependent on the constant
background signal. For a zero-order recovery, $\Delta=\delta$, it is equivalent to a model
of bacterial chemotactic receptors \cite{BarkaiLeibler97}. This system also exhibits exponential adaptive response,
but its steady-state level is independent of background input, a property known as precise adaptation.
It was suggested that this model is related to integral feedback control which implements the maintenance of
output at a constant value \cite{Yi00}.

Recent theoretic work has devoted much attention to the special case of precise adaptive response, in particular in the context of 3-state models which provide the simplest networks to exhibit such a response \cite{Csikasz08,Francois08,MaTang09}. While a circuit that implements precise adaptive response is well suited for homeostatic regulation of its own output, the relation of such a response to the system's sensitivity and dynamic range is often mentioned but rarely quantified. In the following sections we use the model presented here and its extensions to shed light on this relation.

\section{Adaptive enlargement of dynamic range}
\label{sec:DynRange_BasicModel}
We have seen that two functional protein states can define an input/output function $p(u)$ by the balance between them as a function of input signal. Different proteins define different input/output functions depending on the mechanism of interaction with the input. Whatever the exact form of $p(u)$, it has a limited dynamic range that allows a response resolution only within a fraction of the range of possible input signals.

A possible engineering solution to this problem, implemented in many man-made control systems, is a feedback loop which integrates past output,
feeds it back and \emph{subtracts} it from the input. Such feedback shifts the input/output curve on the axis of the input signal and allows its
high-gain region to move \cite{Ogata}; it effectively subtracts constant or slowly varying backgrounds and retains sensitivity to fluctuations
around it. However, in the biochemical systems described by the above models, the feedback loop integrates past activity and feeds it \emph{multiplicatively} onto the input~\cite{FriedlanderBrenner09}. This form of feedback does not shift the response function but only rescales its magnitude. Therefore it cannot compensate for constant backgrounds or enlarge the dynamic range for response. To make these statements more quantitative, we use the model to calculate the response to step inputs $\Delta u$ that appear on top of constant backgrounds $u_0$.

\begin{figure}[]
\centering
\subfigure[]{
\includegraphics[width=0.45\textwidth]{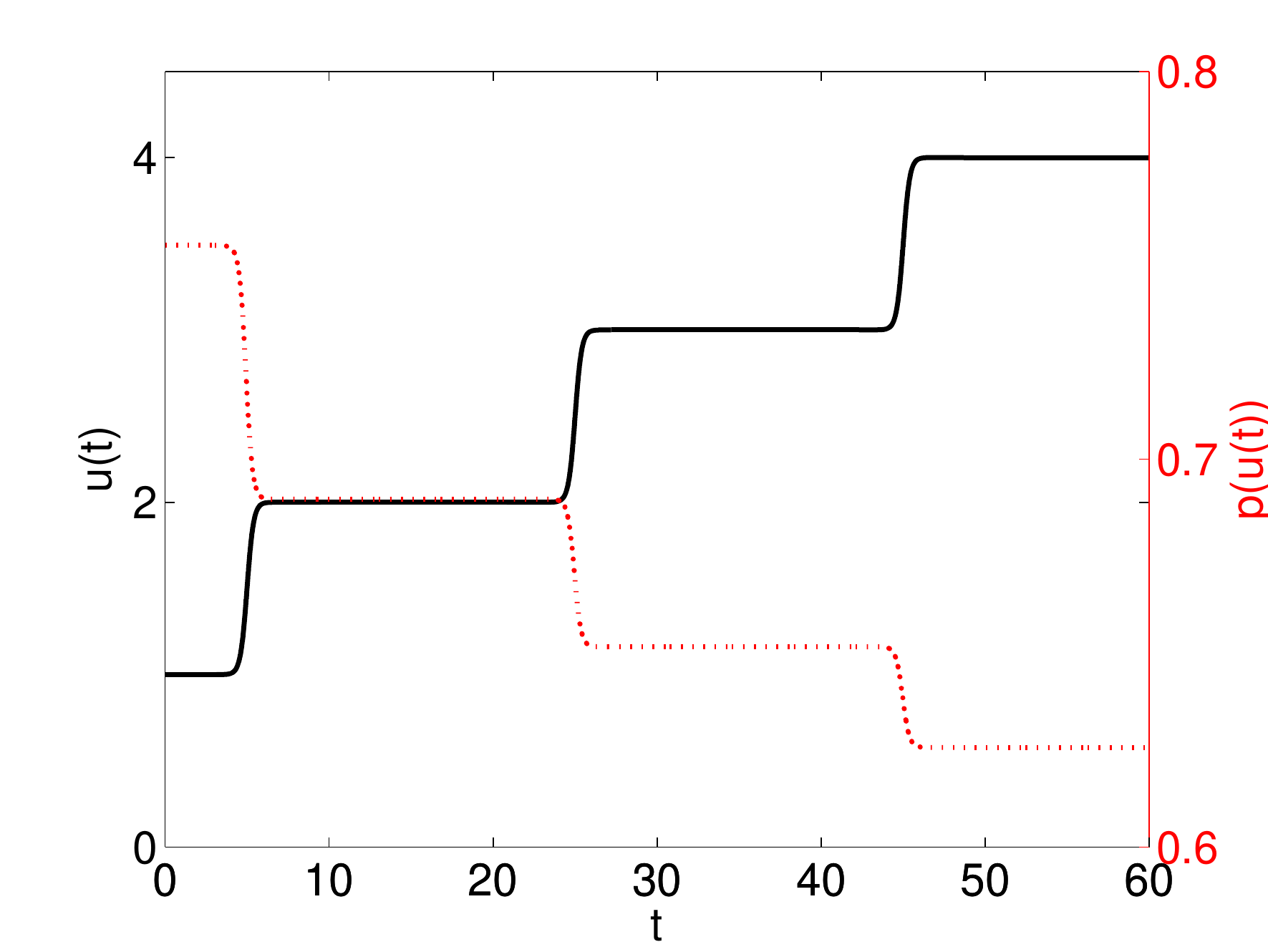}
\label{fig:3_steps_resIn} }
\subfigure[]{
\includegraphics[width=0.45\textwidth]{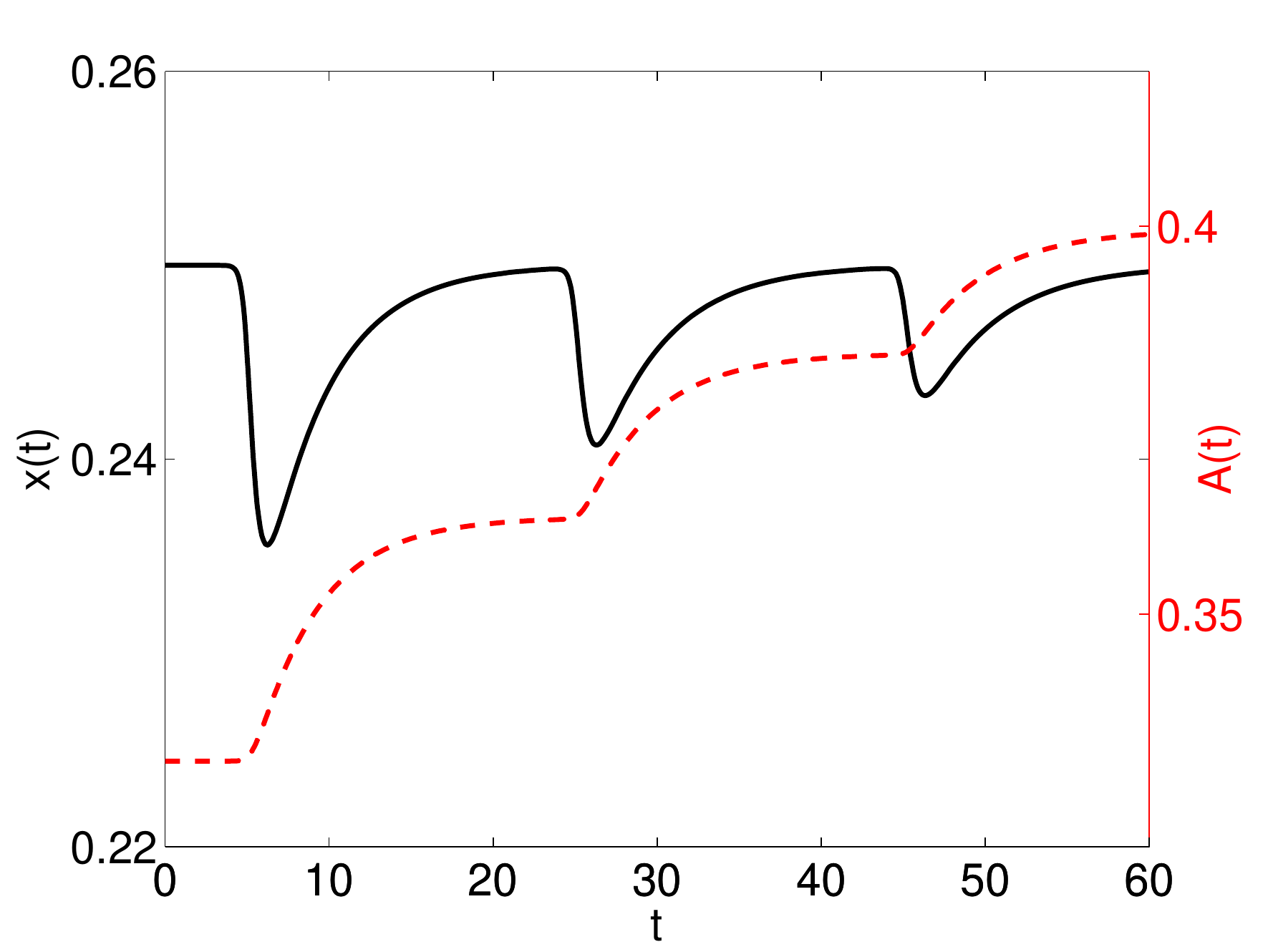}
\label{fig:3_steps_resOut} }
\subfigure[]{
\includegraphics[width=0.45\textwidth]{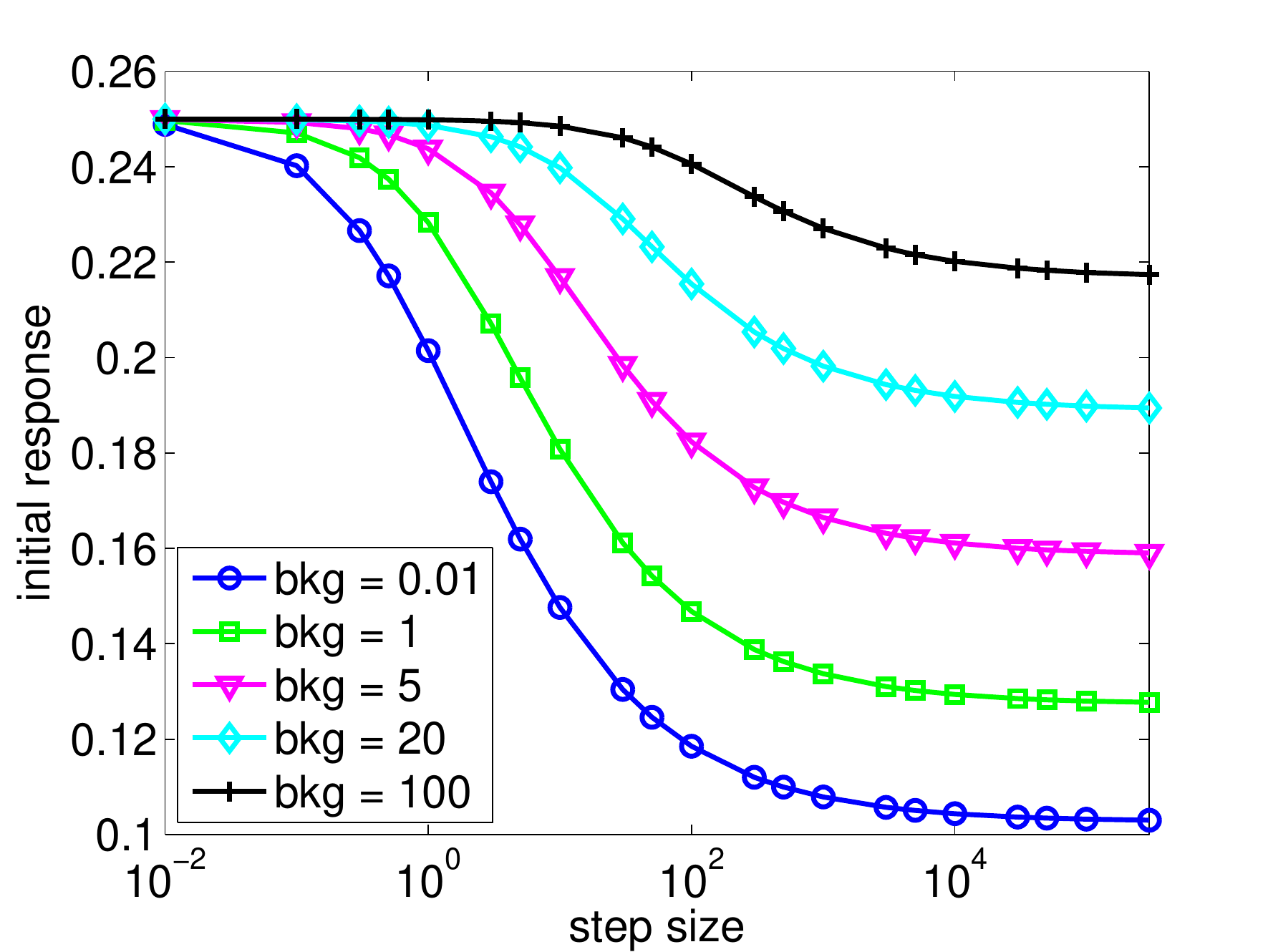}
\label{fig:3_state_exact_p_u_bounded} }
  \caption
 { \label{fig:3_steps_res}
 Dynamic range limitations within a simple 3-state model.
 \subref{fig:3_steps_resIn}: Step input signal $u(t)$ (black line) and
 instantaneous response $p(u(t))$ (red dotted curve).
 \subref{fig:3_steps_resOut}: Activity (black solid curve) and availability (red dashed curve) variables in response
to the input signal shown in \subref{fig:3_steps_resIn}, computed by numeric solution of the dynamical equations
with $\delta = 0.1$, $\gamma = 0.4$, $\alpha(u) = 0.625 + \frac{1}{\sqrt{u+0.1}}$, $\beta(u) = \sqrt{u+0.1}/(1+\sqrt{u+0.1})$;
$p(u) = \alpha(u)/(\alpha(u)+\beta(u))$.
 \subref{fig:3_state_exact_p_u_bounded}: Initial (transient) responses to step inputs $\Delta u$ of different
magnitudes when the system is adapted to various background levels $u_0$.
 }
  \end{figure}

Consider the three-state model introduced in the previous section, for the particular case of zero order recovery,
resulting in a precise adaptive response. This model is
equivalent to the toy model for an adaptive module proposed by Barkai and Leibler \cite{BarkaiLeibler97} and later
studied and extended by many others in the context of bacterial chemotaxis.
Assuming a separation of time scales between the rapid input-dependent transitions and the slow inactivation \eqref{eq:adiab}, one can solve the response dynamics of
\eqref{eq:2D_diff_eq} with $\Delta=\delta$ for an arbitrary input
signal $u(t)$. For a step input,
\be
\label{eq:step}
u(t)=\left\{
\begin{array}{lr}u_0
& t<0
\\
\\
u_0+\Delta u  & t\geq0,
\end{array}
\right.
\ee
the leading term of this approximation will yield:
\be
\label{eq:exact_step_input}
x(t)=\left\{
\begin{array}{lr}\frac{\delta }{\gamma }
& t<0
\\
\\
\frac{\delta }{\gamma}+ \frac{\delta
}{\gamma}\left(\frac{p(u_0+\Delta u)-p(u_0)}{p(u_0)}\right)e^{-t/\tau}  & t\geq0
\end{array}
\right.
\ee
\noindent where
$\tau=1/(\gamma p(u_0+\Delta u))$. 
The infinite-time response here $x^{\infty} := \lim_{t\rightarrow\infty}x=\delta/\gamma$ is independent of the input, the defining feature of precise adaptive response, and the relaxation is exponential.
The magnitude of the transient in this expression shows manifestly how the nonlinear response function
$p(u)$ decreases the response magnitude to a step $\Delta u$ as the background $u_0$ increases.
A graphical illustration of this result for several step functions is shown in \figref{fig:3_steps_res} \footnote{Throughout this paper we have used response functions $p(u)$ that are decreasing functions of the input signal $u$, to comply with the
conventional definition of receptor activity in bacterial chemotaxis.}.

To quantify the ability of a system to enlarge its dynamic range by background compensation,
one can use such step-function inputs
as follows: for each ambient background signal $u_0$, the system is probed by increasing steps of
input $\Delta u$ on top of this background. For each background a curve is then plotted as a function of the step
magnitude. An effective enlargement of dynamic range occurs when the different curves shift horizontally on the
input axis, such that their sensitive, unsaturated portion is centered around the constant background.
\figref{fig:3_state_exact_p_u_bounded} shows this plot for the three-state model with precise adaptive response
considered here. The response curve is not shifted horizontally but remains in the same vicinity of input while
significantly decreasing in amplitude.
We conclude that an adaptive response in system activity, even if precise, is insufficient for an adaptive
enlargement of dynamic range by background compensation; moreover it is insufficient to maintain the entire
dynamic range of system output, although it returns to the same output value at steady state
regardless of the constant background (this is reflected by all curves in the figure starting at the same output
value at $\Delta u =0$).
In fact, none of the three-state models represented by the general class \eqref{eq:COI_scheme}
shows the property of dynamic range enlargement. In the following section we show a generalization
of the available/unavailable states to a graded ladder of responsiveness,
that allows a population of receptors to implement a weighted sum of several input/output functions and
thus to shift and rescale the range of input to which it effectively responds.

\section{Graded responsiveness as a mechanism for adaptive dynamic range enlargement}
\label{sec:ExtModel}

\begin{figure}[]
\begin{center}
  \includegraphics[width=0.9\textwidth]{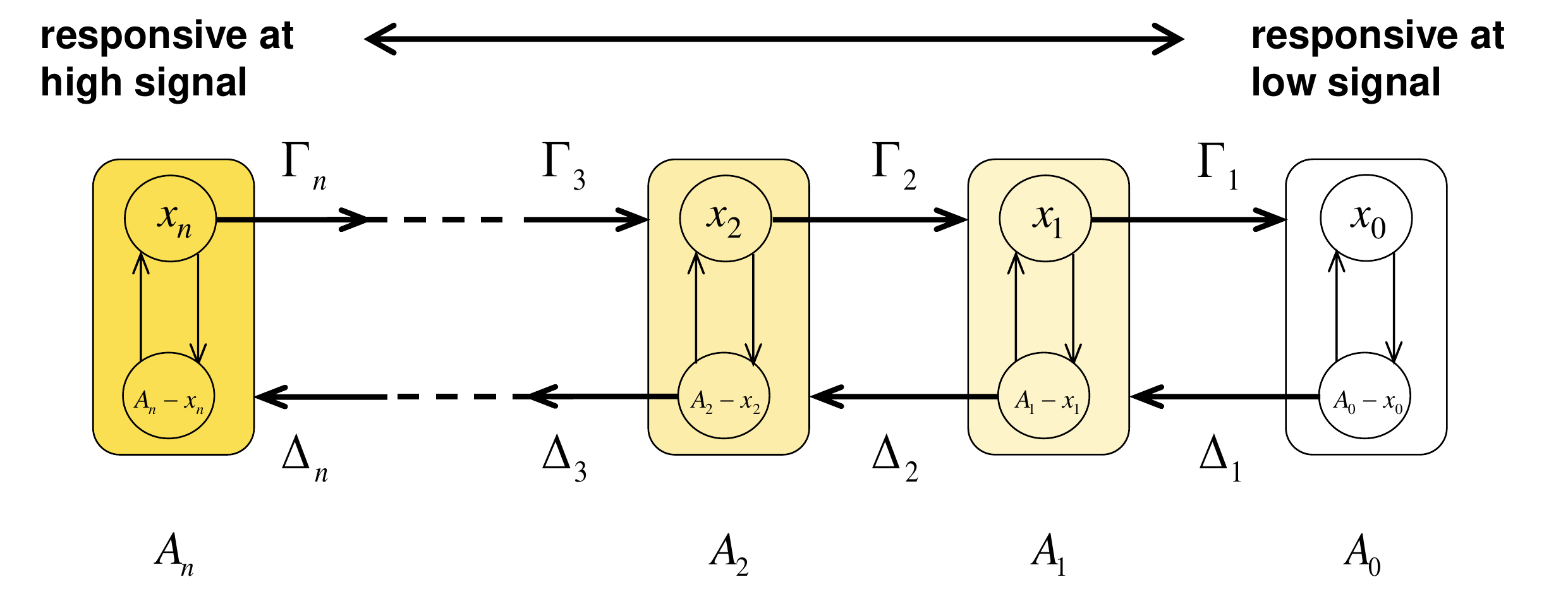}\\
  \caption{Illustration of the graded responsiveness model. Proteins can occupy multiple responsiveness classes $A_0$ to $A_n$; in each class there are both active and inactive proteins
  rapidly equilibrating in response to the input signal. The equilibrium fractions determine the class's characteristic response function $p_i(u)$. Transitions between classes are slow and activity-dependent, and are
  represented here by the general terms $\Gamma_i$ and $\Delta_i$.
}\label{fig:GradedAvail}
  \end{center}
\end{figure}

In the simplified class of models discussed above (\eqref{eq:COI_scheme}),
we considered a division of the ensemble of
molecules into two classes: available or unavailable to respond to input signal. This is a coarse-grained picture of the molecular system which simplifies the model and is often invoked for simplicity of analysis. The resulting three-state models exhibit adaptive response which, if precise, can serve as an effective homeostasis mechanism. However we have shown that this property does not reflect an enlargement of the dynamic range for response to signals. Here we show that a more gradual modulation of the
molecule into classes of varying sensitivity, can account for an effective background compensation over a broad range of inputs, such as observed in experiments on several signaling systems \cite{Dunten91}.

In bacterial chemotaxis, it is well established that multiple methylations play a central role in adaptation to background. Receptors have several methylation sites
and adaptation involves a change in the average methylation level per receptor \cite{Terwilliger86}. Asakura and Honda \cite{Asakura84} constructed a multi-state model relying on the properties of methylation in bacterial receptors.
In order to account for experimental results they constrained the model parameters so as to obtain a precise adaptive response.

We present below a generalized version of the mechanism they proposed, and analyze its properties in terms of adaptive response and dynamic range. Our contribution to previous results will be in quantifying the relations between the kinetic details of the model and the emerging system properties, namely adaptive response and dynamic range. The main conclusion will be that this model enlarges dynamic range regardless of the type of adaptive response it induces. Relieving this constraint makes the mechanism more generally applicable; we discuss such possible applications later.

Imagine a set of protein states, physically modified one with respect to another; this is displayed by the scheme in \figref{fig:GradedAvail} with $n+1$ different classes,
occupied by relative concentrations $A_0$ to $A_n$ $(\sum_{i=0}^n A_i=1)$. Within each class there are both an active (occupied by a concentration $x_i$) and an inactive (with $A_i-x_i$)
state. Each class is characterized by a different degree of responsiveness to the input signal, reflected by a shifted balance between the active and inactive states at a given value of input. Thus, as in the simpler model, we distinguish two types of labels on the molecular states: active and inactive are functionally distinct, whereas responsiveness class $i$
includes both these states but exhibits a different affinity towards the input signal.
This is a generalization of the availability concept presented in the basic 3-state model (\eqref{eq:COI_scheme} and \figref{fig:3-state}) incorporating only two classes, one of which was completely non-responsive to the signal. To reflect this analogy we keep the notation similar, with class occupancies denoted by $A_i$ and class decrease and increase transitions by $\Gamma_i$ and $\Delta_i$, correspondingly.

The responsiveness to input is characterized, as before, by an instantaneous
 input/output function for each class, $p_i(u)$. Indeed measurements on chemotactic receptors with fixed methylation levels \cite{LevitStock02}, have shown that the input/output functions are shifted one with respect to the other, covering different regions of input signal by their sensitive parts. In particular the least methylated
receptors are first to saturate when the
signal increases, and the most methylated saturate last \cite{Dunten91,LevitStock02}. Therefore this type of state space provides a degree of freedom for the ensemble of receptors to modulate its response by redistribution among the modified states.

\begin{figure}[]
\centering
\subfigure[]{
\includegraphics[width=0.45\textwidth]{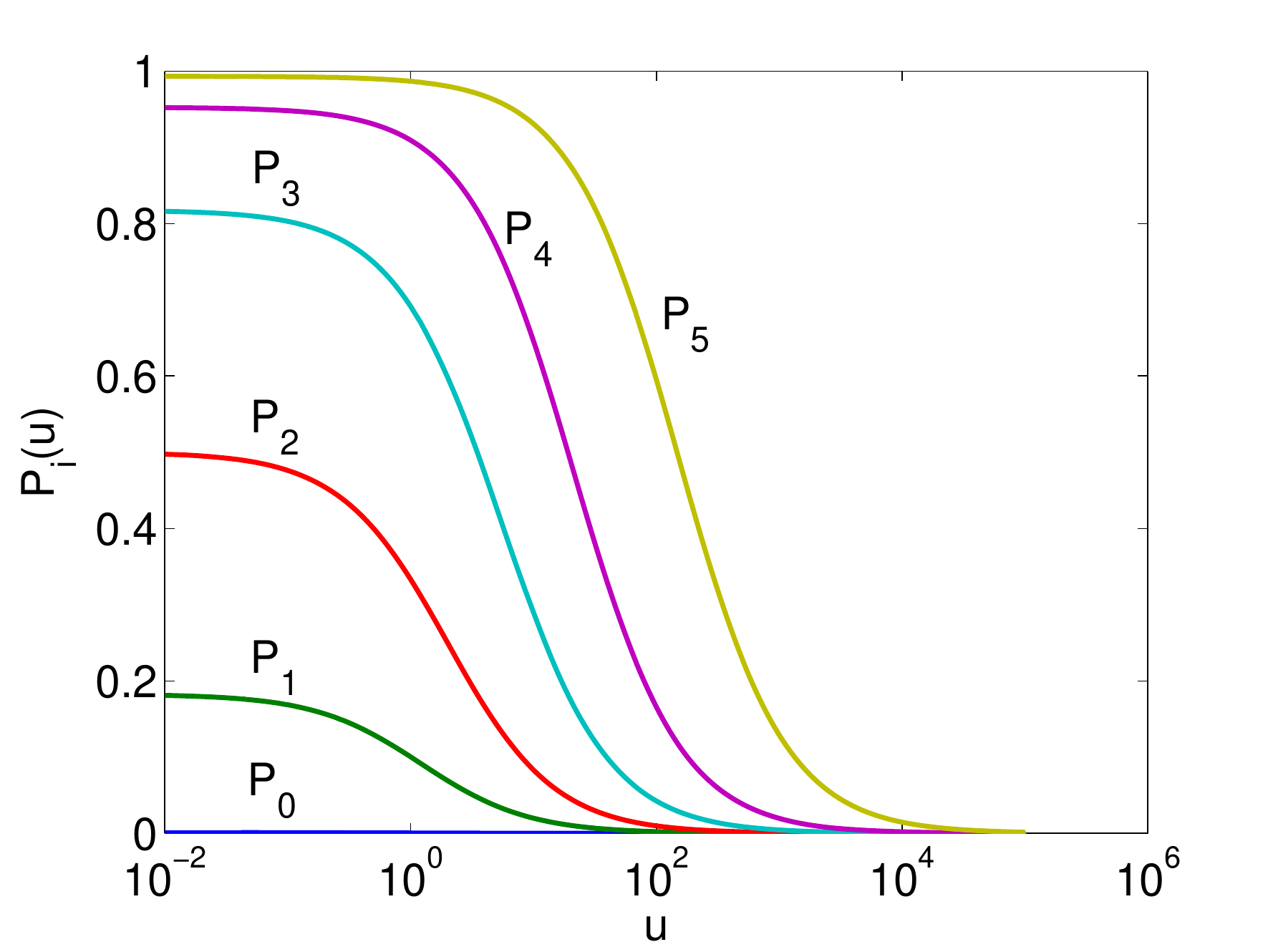}
\label{fig:p_i_of_u} }
\subfigure[]{
\includegraphics[width=0.45\textwidth]{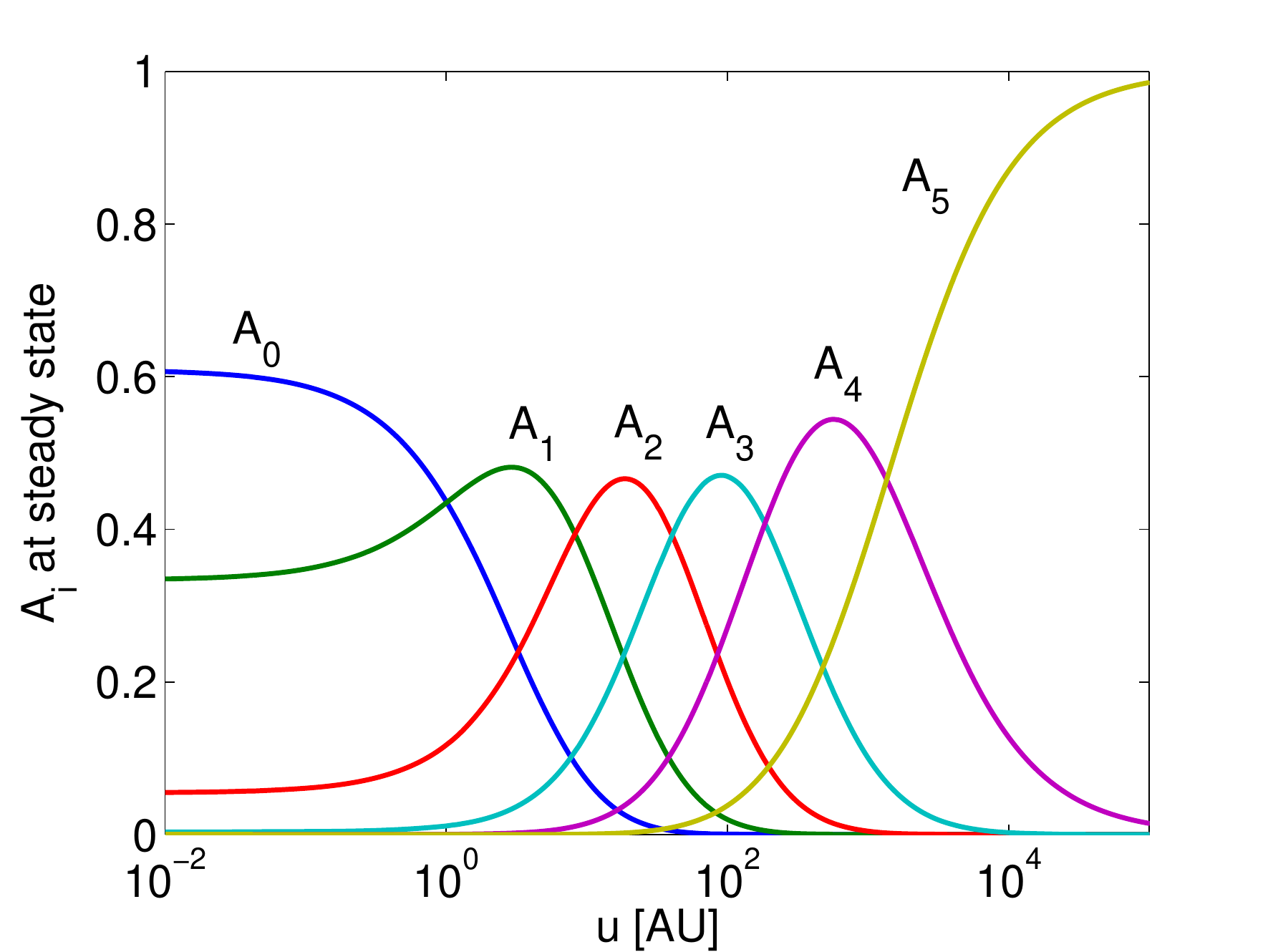}
\label{fig:MethylDistSS} }
  \caption
 { \label{fig:GradedMethyl}
 Dynamic range enlargement is a population effect.
 \subref{fig:p_i_of_u} Input/output functions $p_i(u)$ of the various classes. Here
  $p_i(u)=1/(1+e^{K_i}\!\cdot\!(1+u))$, with $K_i=7,1.5,0,-1.5,-3,-5$.
 \subref{fig:MethylDistSS} Distribution of availability class occupancy by receptors at steady state vs. the signal level in the graded responsiveness model of \figref{fig:GradedAvail}.
 Adaptation involves changes in class distribution (see also Fig. 1 in \cite{Asakura84}).
In this model, only active receptors can be de-modified and only inactive ones can be modified.
}
  \end{figure}

The weighted sum $x=\sum x_i = \sum p_i(u) A_i$ represents the system total activity and is assumed to be directly related to its output to downstream processes. Adaptive responses are usually referred to and measured in terms of this output.
In addition to this measurable quantity, the state of the system is also characterized by the underlying distribution of receptors among responsiveness classes, $\{A_i\}_{i=0}^{n}$ -
see numerical example in \figref{fig:MethylDistSS}.
The mean modification level per receptor is the average of this distribution:
$\mathcal{A}\equiv\displaystyle\sum_{i=1}^n i A_i$.

Transitions between responsiveness classes, here represented by the arbitrary terms $\Gamma_i$ and $\Delta_i$, are slow and activity dependent:

\begin{equation}
\label{state_scheme}
A_n\mathop{\rightleftarrows}^{\Gamma_n}_{\Delta_n} \ldots
\mathop{\rightleftarrows}^{\Gamma_3}_{\Delta_3} A_2
\mathop{\rightleftarrows}^{\Gamma_2}_{\Delta_2} A_1
\mathop{\rightleftarrows}^{\Gamma_1}_{\Delta_1} A_0
\end{equation}

\begin{figure}[]
\centering
\subfigure[]{
\includegraphics[width=0.45\textwidth]{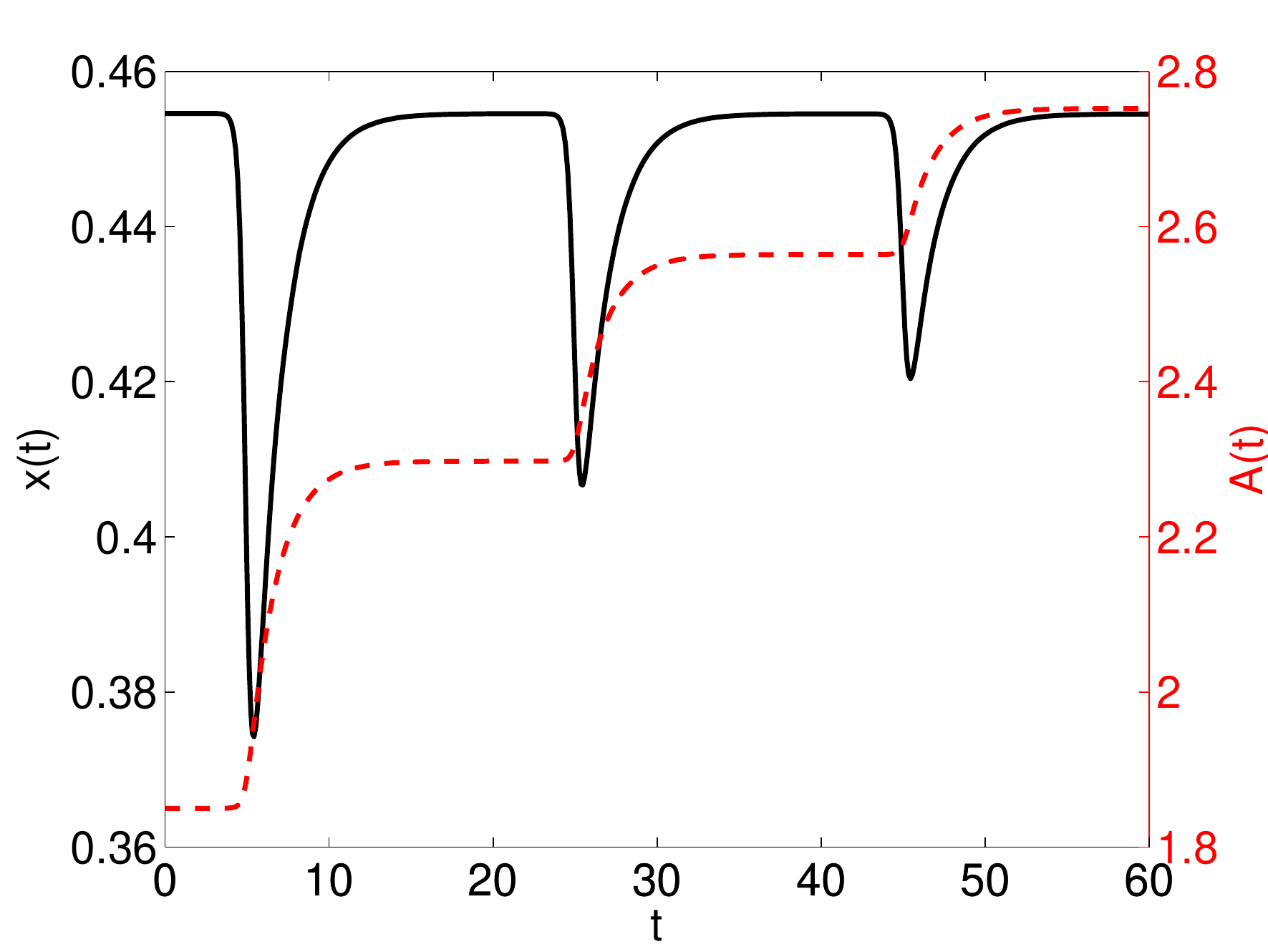}
\label{fig:TotAct_ExtAvail} }
\subfigure[]{
\includegraphics[width=0.45\textwidth]{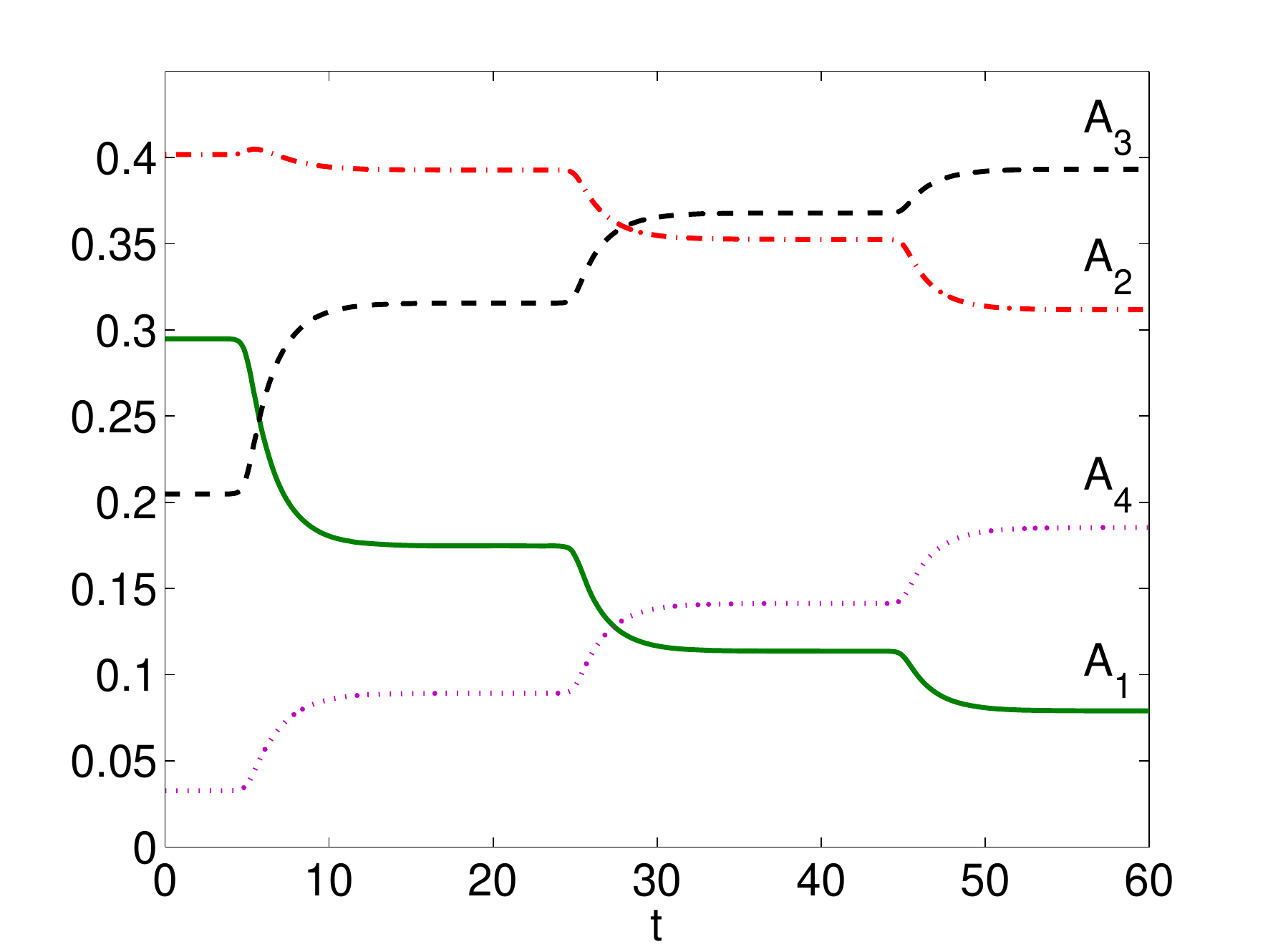}
\label{fig:GradedAvail_3_steps_res} }
  \caption
 { \label{fig:GradedAvailRes} Response to step signals in the graded availability model.
 \subref{fig:TotAct_ExtAvail}: Changes in total activity $x(t)$ and mean modification level $\mathcal A(t)$
in response to 3 steps of magnitude 1.
 \subref{fig:GradedAvail_3_steps_res} Occupancy of classes $A_1$-$A_4$ under the same input.
Here the input/output functions $p_i(u)$ are as in \figref{fig:GradedMethyl}.
with $K_i=7,1.5,0,-1.5,-3,-7$; Transitions betwee classes are first order $\Delta_i=A_i-x_i, ~~\Gamma_i=1.2x_i$.
 }
  \end{figure}

\noindent
This activity dependence provides the necessary feedback from the system output to its effective response function.
\figref{fig:TotAct_ExtAvail} shows an example of the total activity $x$ in response to steps, in this case exhibiting a precise adaptive response. \figref{fig:GradedAvail_3_steps_res} shows that the redistribution among classes is altered, although system activity returns to the same value it had before the
additional step input.

The adaptive response in system activity is a result of the existence of two separated timescales in the system, causing two stages of response to a change in input. Equilibration
between active and inactive states happens rapidly with the change in signal, whereas the redistribution among responsiveness classes is a slower process that results in the later
relaxation in activity. This is a property of the general class of
models depicted in \figref{fig:GradedAvail}, irrespective of the detailed kinetics of transitions.
However, in order for the system activity to display precise adaptive response,
special assumptions need to be made on the kinetics of transitions between responsiveness classes.
Different assumptions were made by different authors in order to constrain the model to display this precise adaptive
response (see Appendix).
However, the structure defined by \figref{fig:GradedAvail} exhibits an adaptive dynamic range enlargement well beyond
these constraints, for various types of kinetics defined by the abstract symbols $\Gamma$ and $\Delta$.
Therefore adaptive dynamic range enlargement can be found in such a model independent of the precision of adaptive response.
For example, if the kinetics of modification and de-modification is first order with arbitrary rate constants,
the adaptive response is imprecise (steady state is input-dependent), but nevertheless the system maintains its sensitivity
to input changes on top of large backgrounds. \figref{fig:StepResDiffBkgAdpt} shows the quantitative summary of these
statements for the graded-responsiveness model with precise (\figref{fig:DiffBkgAdpt_exact}) and imprecise
(\figref{fig:DiffBkgAdpt_non_exact}) adaptive response. As expected, the dynamic range of response is slightly diminished
by the imprecision of the steady state, since the range of activity (output) is itself limited (range of curve in the y-axis).
However the sensitivity towards input fluctuations on top of a background is broadened in a similar manner in both cases,
demonstrating a logarithmic horizontal shift of the response curves with the signal (x-axis). We recall that such behavior is
not exhibited by the 3-state model (compare to
\figref{fig:3_state_exact_p_u_bounded}).

\section{Discussion}
\label{sec:Disc}

Adaptive response and high sensitivity over an extended dynamic range are properties of many biological signaling
systems. In addition to developing realistic models describing the specific aspects of each system,
it is of interest to study on a fundamental level which model ingredient gives rise to what system property.
In this work we have used abstract simplified models to disentangle adaptive response properties from
signal-processing properties, namely the ability to compensate for constant background and maintain sensitivity
to transients over a broad range of inputs.

The characteristic feature of background compensation is a shift of the transient-response function along
the input axis as the background changes. This shift allows the sensitive portion of the input/output function
to be centered where fluctuations in signal are expected to occur; it is generally implemented by internal
degrees of freedom that are not directly reflected by the system output.
On the other hand, adaptive response is a dynamic property of system output.
It is not surprising therefore that the relation between these two properties is not one-to-one.

We demonstrated that adaptive response alone is insufficient to provide an enlarged dynamic range,
even if it is precise. By analysing the simplest biochemical models displaying adaptive response we have
shown in our previous work that the circuit
implemented by these systems contains a multiplicative, rather than additive, feedback branch
\cite{FriedlanderBrenner09}. Such feedback can induce a constant (input-independent) steady-state output,
namely a precise adaptive response. Here we have shown explicitly that this property does not ensure availability
of the entire range
of output for further stimulation. More importantly, it does not induce an adaptive shift of the input/output
response curve
on the input axis, and thus does not allow its sensitive portion to move to different regions of input.
Our conclusion is that
these simplified models display only a phenomenological dynamic effect of adaptive response,
which does not necessarily fulfill any functional role in signal processing.

Additional internal degrees of freedom are
required to induce the flexibility of adaptive background compensation.
One mechanism that has been proposed \cite{Asakura84} is the modulation of the protein input/output response
by multiple modifications, with each modification shifting the sensitive region around different input values.
Changes in the distribution of proteins among the different modification classes can then
effectively change the total response of the system. Quantitatively, the extent and nature of the
response shift will be dictated by the input/output response curves at the different modification classes.
On a coarse-grained level of description, the change in distribution can be approximated by the change in average
modification level of the receptor population. This can be described by a continuous
modulating parameter of the average input/output response curve of the entire population \cite{TuBerg08}.

Building on this previous model for bacterial chemotactic receptors \cite{Asakura84},
we have analyzed a general structure of protein states which induces the property of dynamic-range enlargement.
We have shown that this mechanism maintains sensitivity to changes on top of background within a broad range
of input signals, independent of several kinetic details and regardless of the type of adaptive response it
induces. Precise adaptive response can expand the range of transient output to cover the entire available range
of system activity, namely affect the amplitude of response. However, the horizontal shift representing the
background compensation is a property of a much broader class of models that displays arbitrary adaptive response.

\begin{figure}[]
\centering
\subfigure[]{
\includegraphics[width=0.45\textwidth]{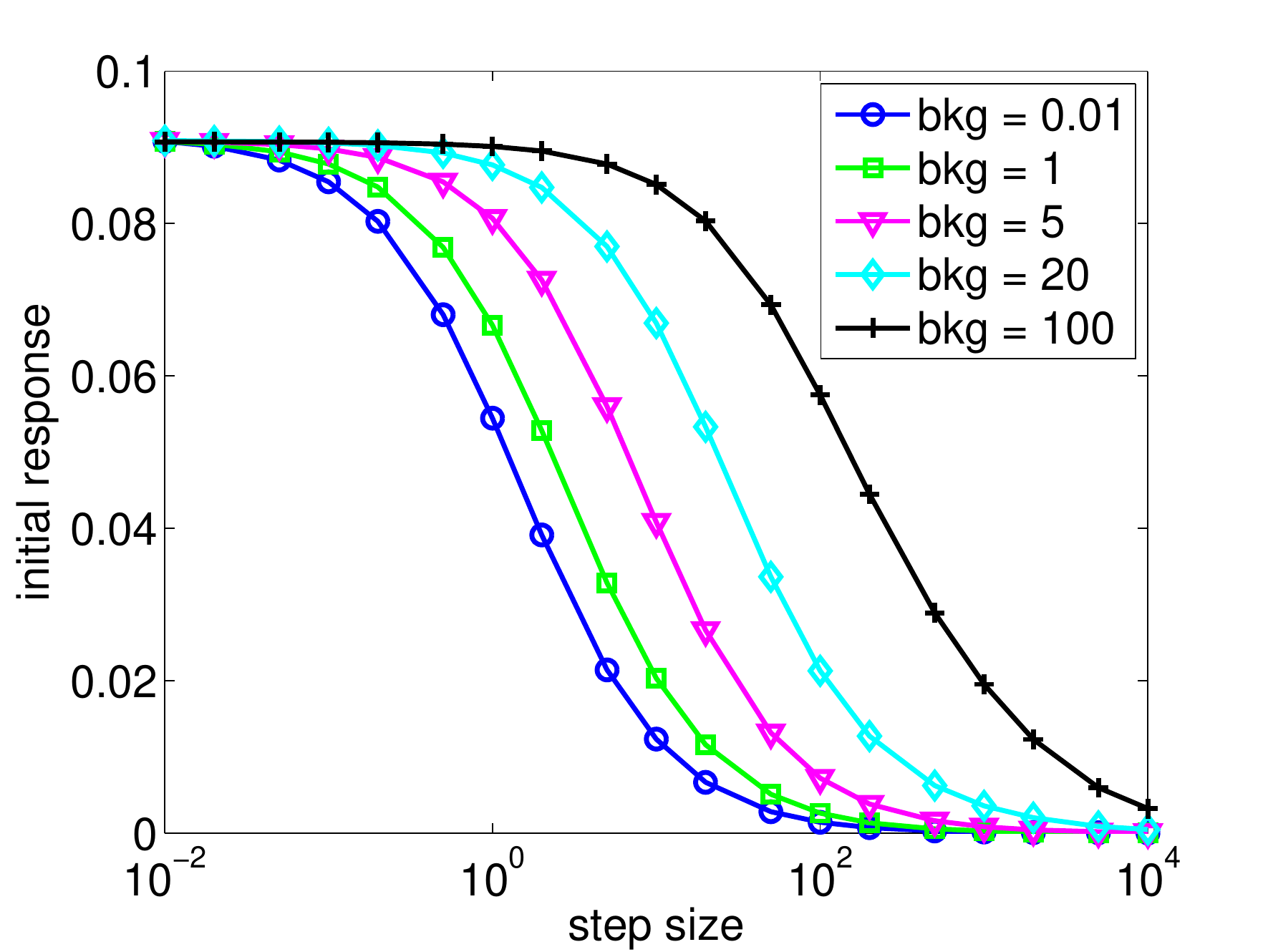}
\label{fig:DiffBkgAdpt_exact}}
\subfigure[]{
\includegraphics[width=0.45\textwidth]{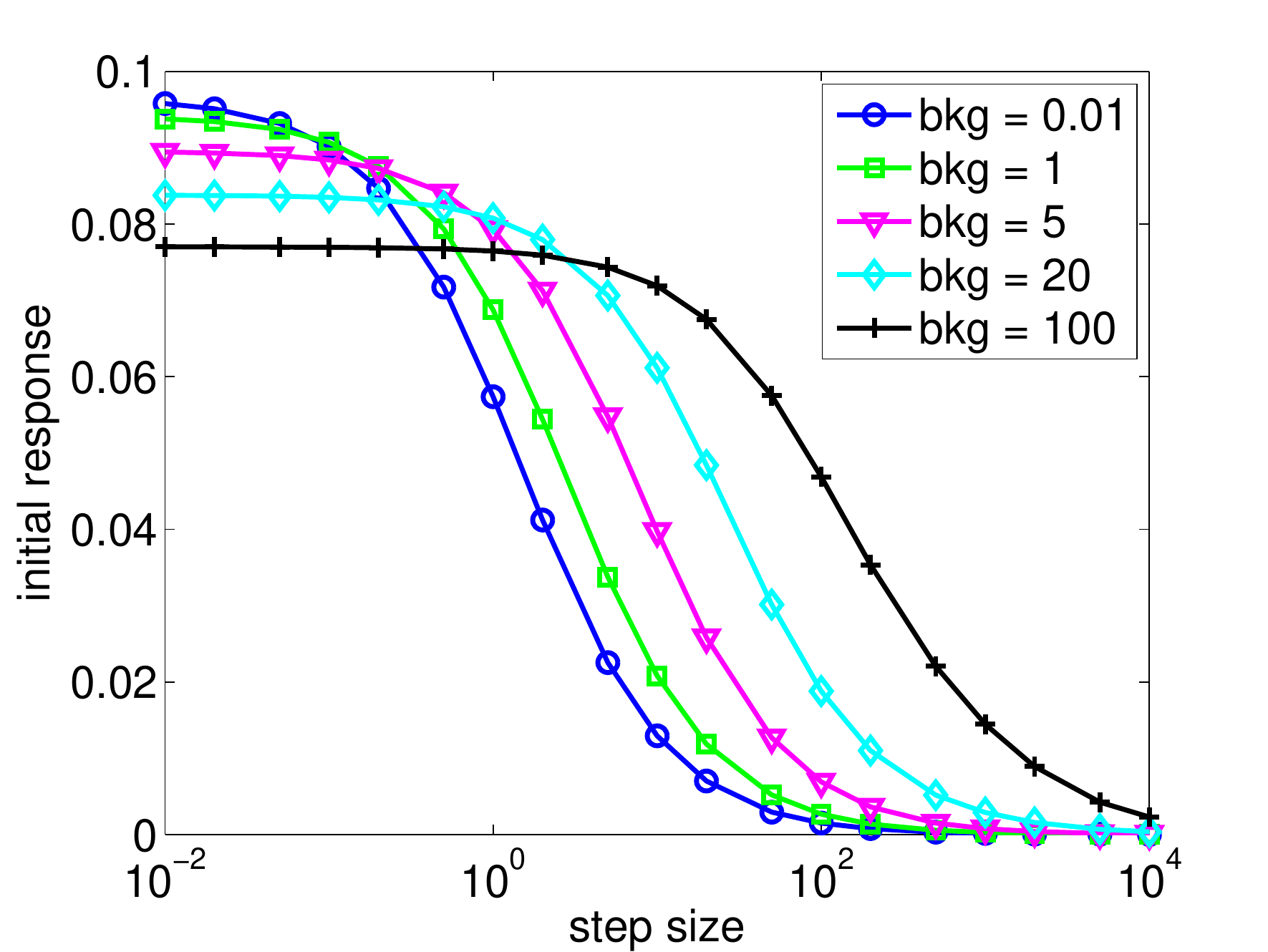}
\label{fig:DiffBkgAdpt_non_exact}}
  \caption
 { \label{fig:StepResDiffBkgAdpt}
Initial (transient) responses to step inputs of different magnitudes, when the system is adapted to various background levels in the 6-class graded responsiveness model.
\subref{fig:DiffBkgAdpt_exact} Parameters inducing precise adaptive response; same as in \figref{fig:GradedAvailRes}.
\subref{fig:DiffBkgAdpt_non_exact} Parameters inducing imprecise adaptive response; first-order transitions, $\Delta_i=\delta_i\cdot(A_i-x_i)$ and $\Gamma_i=\gamma_i x_i$ with arbitrary rate constants
$\delta_1=1.1$, $\delta_2=1$,
 $\delta_3=0.9$, $\delta_4=0.83$, $\delta_5=0.75$, $\gamma_i= 10$.
 }
  \end{figure}

In bacterial chemotactic receptors, multiple methylation of the receptor provides the ladder of modification,
and recent studies are starting to unravel the molecular mechanisms underlying graded responsiveness in these
receptors \cite{Hazelbauer2008}.
Other processes such as multiple phosphorylation can in principle induce similar properties in other signaling
proteins.
Indeed it has been suggested that in photoreceptors multiple phosphorylation plays an analogous role to
methylation \cite{Terwilliger86,Dunten91}.
The results presented in \figref{fig:StepResDiffBkgAdpt} are remarkably similar to experimental results on
bacterial chemotaxis (Fig. 3A of \cite{Sourjik02}) and on photoreceptors
(Figs. 7 and 9 of \cite{NormannPerlman79}), supporting the generality of this mechanism in widely diverse
sensory systems (bacterium cell vs. turtle's photoreceptor cell). Further
experiments are needed to test this idea and to explore the degree of universality between the mechanisms.

\section*{Appendix: Precise adaptive response in the graded-responsiveness model}
In this appendix we consider constraints on the kinetics of the general model displayed in \figref{fig:GradedAvail} such that precise adaptive response ensues.
Asakura and Honda \cite{Asakura84} in their original model proposed the following conditions: 
(1) both class-increase (methylation) and class-decrease (de-methylation) reactions are state-dependent, i.e. de-methylation works only on active receptors and methylation works only on inactive ones. (2) the ratio between methylation and de-methylation transition rates are independent of methylation level; i.e. $\Delta_i/\Gamma_i$ is independent of $i$. (3) the end states $A_0$ and $A_n$ are
characterized by a very extreme equilibrium such that each of them includes effectively only one activity state. Under these assumptions a precise adaptive response is observed over a
range of parameters. Barkai and Leibler \cite{BarkaiLeibler97}, on the other hand, attributed the appearance of precise adaptive response to the action of the de-methylation enzyme CheR at saturation \cite{Lupas89, Simms87} such that the total rate of this reaction is fixed. An application of this assumption to the current model is presented below. In either case the requirement for precise adaptive response, such as is observed over a range of parameters in bacterial chemotaxis experiments, places constraints on the kinetics of the model.

We have seen that within the simplified binary model of state-dependent inactivation, a zero-order kinetics of return from unavailability results in exponential, precise
adaptive response. 
Now in the graded-responsiveness generalization, assuming the de-modifying enzyme acts independently of the current modification level, the saturation condition implies $\sum_{i=1}^{n}
\Delta_i=\delta$. Activity-dependent kinetics enters here by the de-modification reaction acting only on active receptors,
hence depending on the concentrations of active molecules $x_i$. Assuming that this
reaction is first-order and also insensitive to modification level, it depends on the total concentration
of active molecules: $\sum_{i=1}^{n} \Gamma_i = \gamma \sum_{i=1}^{n} x_i\approx
\gamma\sum_{i=0}^{n}x_i=\gamma x$. Under these assumptions all transitions in this direction depend on the total activity; relaxing it implies that it depends on a more complex weighted average of activities at different classes.

In the adiabatic approximation $x_i$ equilibrates rapidly in
response to a change in the signal such that $x_i\approx p_i(u)A_i$.
The slower
variables $A_i$ then obey the following equations:
\begin{eqnarray}
\label{eq:A_i} \dot{A_0}&=&-\Delta_1 + \Gamma_1 \\\nonumber \dot{A_1}&=&+\Delta_1
-\Gamma_1 - \Delta_2 + \Gamma_2\\\nonumber \dot{A_2}&=&+\Delta_2 - \Gamma_2 - \Delta_3 +
\Gamma_3\\\nonumber &\vdots&
\\\nonumber \dot{A_n}&=&+\Delta_n - \Gamma_n\nonumber
\end{eqnarray}

\noindent whereas the mean modification level obeys
\begin{eqnarray}
\dot{\mathcal{A}}&=& \sum_i i \dot{A}_i \\\nonumber &=&\Delta_1 -\Gamma_1 - \Delta_2 +
\Gamma_2 +2(\Delta_2 - \Gamma_2 - \Delta_3 + \Gamma_3)+\ldots +n(\Delta_n -
\Gamma_n)\\\nonumber &=&\sum_{i=1}^{n} \Delta_i -\sum_{i=1}^{n} \Gamma_i\nonumber.
\end{eqnarray}

These assumptions imply that the mean modification retains the same relation with the total system activity as the availability $A$ in the three-state model \cite{FriedlanderBrenner09}:
\be \dot{\mathcal{A}}=-\gamma x + \delta,\ee
showing manifestly that the system's steady state response is independent of the input stimulus
\be x^\infty=\sum_i x_i^\infty = \sum_i p_i(u)A_i = \frac{\delta}{\gamma}\ee

\noindent compare this result to \cite{Yi00}.


\section*{Acknowledgments}
We are grateful to Shimon Marom, Erez Braun and Ron Meir for fruitful discussions related to this article.

\bibliographystyle{unsrt}
\bibliography{MyBib}

\end{document}